\documentclass[preprint,authoryear,12pt]{elsarticle}%
\usepackage{amssymb}
\usepackage{graphicx}
\usepackage{epsfig}
\usepackage{amsthm}
\usepackage{longtable}
\usepackage{lscape}
\usepackage{amsmath}
\journal{Journal of High Energy Astrophysics}

\begin{document}

\begin{frontmatter}



\title{Testing Einstein's Equivalence Principle with Supercluster Laniakea's Gravitational Field}


\author[label1]{Zhi-Xing Luo}
\author[label1]{Bo Zhang}
\author[label1]{Jun-Jie Wei}
\author[label1,label2]{Xue-Feng Wu\corref{dip}}
\ead{xfwu@pmo.ac.cn}

\address[label1]{Purple Mountain Observatory, Chinese Academy of Sciences, Nanjing 210008, China}
\address[label2]{Joint Center for Particle, Nuclear Physics and Cosmology, Nanjing
University-Purple Mountain Observatory, Nanjing 210008, China}
\cortext[dip]{Corresponding author.}

\begin{abstract}
Comparing the parameterized post-Newtonian parameter $\gamma$ values for different types of particles, or the same type of particles with different energies is an important method to test the Einstein Equivalence Principle (EEP). Assuming that the observed time delays are dominated by the gravitational potential of the Laniakea supercluster of galaxies, better results of EEP constraints can be obtained. In this paper, we apply photons from three kinds of cosmic transients, including TeV blazars, gamma-ray bursts as well as fast radio bursts to constrain EEP. With a gravitational field far more stronger than a single galaxy, we obtain 4--5 orders of magnitude more stringent than the pervious results.
\end{abstract}


\end{frontmatter}

% \linenumbers

\def\astrobj#1{#1}
\section{Introduction}
\label{sect:intro}

The Einstein Equivalence Principle (EEP) stands as one of the most important basic assumptions as well as cornerstones of general relativity, along with many other metric theories of gravity. According to EEP, the traveling path of any uncharged test object in vacuum is independent of the object's internal structure and composition. All metric gravity theories taking EEP as assumption predict one of the parameterized post-Newtonian (PPN) parameter $\gamma_{1}=\gamma_{2}\equiv\gamma$, where the subscripts 1 and 2 denote two different test particles (such as photons or neutrinos), respectively
\citep{2006LRR.....9....3W,2014LRR....17....4W}. So the accuracy of the EEP can be constrained by comparing the value of $\gamma$ for different types of particles, or the same type of particles with different energies.

Many methods have been developed to test the EEP with high accuracy by the measurement of the value of $\gamma$. One of the most successful method, which measures the gravitational deflection of light near the Sun and the round-trip travel time delay of artificial radar signal due to the solar system gravity, yields $\gamma-1=(-0.8\pm1.2)\times10^{-4}$ \citep{2009A&A...499..331L,2011A&A...529A..70L} and $\gamma-1=(2.1\pm2.3)\times10^{-5}$ \citep{2003Natur.425..374B}. Recently, EEP has also been tested using the time delay of photons with different energies arising in single cosmic transient event, such as gamma-ray bursts \citep[GRBs;][]{2015ApJ...810..121G}, fast radio bursts \citep[FRBs;][]{2015PhRvL.115z1101W,2016arXiv160207643}, and TeV blazars \citep{2016ApJ...818L...2W}. These results have been improved for several orders of magnitude compared with previous works, that is, $\gamma_{GeV}-\gamma_{MeV}<2\times10^{-8}$ for GRB 090510 \citep{2015ApJ...810..121G}, and $\gamma_{1.23GHz}-\gamma_{1.45GHz}<4.36\times10^{-9}$ for FRB 100704 \citep{2015PhRvL.115z1101W}, and $\gamma_{(0.2TeV-0.8TeV)}-\gamma_{(>0.8TeV)}<2.18\times10^{-6}$ for TeV blazar PKS 2155-304 \citep{2016ApJ...818L...2W}.
Very recently, such EEP tests have also been applied to gravitational waves \citep{2016arXiv160201566W,2016arXiv160204779K}.

In all these works, to account for the time delay of the photons, gravitational field taken into consideration were from Milky Way only. However as mentioned by \citet{2015arXiv160103636N}, \citet{2015arXiv160104558Z}, and \citet{2016arXiv160206805W}, larger scale structures, such as galaxy clusters and other large scale fluctuations have stronger gravitational potentials, thus may cause larger delay between different particles. Such structures can provide even better constraints on EEP. In this work we test the EEP with photons from various astrophysical transients, adopting the Laniakea, the supercluster in which Milky Way resides as our source of gravitational field. Since Laniakea supercluster of galaxies is much more massive than a single galaxy, more stringent constraints can be obtained. In Section 2 of this paper, our basic methods are described. Our results are presented in Section 3. In Section 4, this analysis is discussed and summarized.

\section{Constraining EEP with Gravitational Field of Laniakea}
\label{sect:CSFR}
Superclusters are the most massive structure in cosmic scales. Although each member cluster of galaxies is not affected by mutual gravitational forces, supercluster can influence not only the motions of its member clusters, but also the expansion of the Universe itself. Laniakea is a newly discovered supercluster of galaxies to which the Milky Way galaxy belongs \citep{2014Natur.513...71T}. This supercluster encompasses some 100,000 galaxies in 300 to 500 galaxy clusters, and stretches more than 500 million light-years. The total mass of the Laniakea is $10^{17}$ solar masses, which is nearly a hundred thousand times that of our Milky Way galaxy.

The limits on the differences in PPN parameters for difference particles determine the accuracy of EEP. For example, it has been shown that the time interval required for photons to traverse a given distance is longer in the presence of a gravitational potential $U(r)$ by
\begin{equation}
\Delta t=-\frac{1+\gamma}{c^{3}}\int^{r_{o}}_{r_{e}}U(r)dr,
\end{equation}
where $r_{e}$ and $r_{o}$ are the locations of the emission and observation, respectively \citep{1964PhRvL..13..789S,1988PhRvL..60..173L,1988PhRvL..60..176K}. Here $\gamma$ is one of the PPN parameters, representing the space curvature produced by unit rest mass. $\gamma$ is found to be nearly unity, consistent with the prediction of $\gamma=1$ by general relativity \citep{1993tegp.book.....W}. However, in testing EEP, a more important question is whether different types of particles share the same value of $\gamma$, rather than the absolute value of $\gamma$.

As shown in \citet{2015ApJ...810..121G} and \citet{2015PhRvL.115z1101W}, for a cosmic transient source, the various terms that might contribute to the observed time delay between two different energy bands may be expressed as follows:
\begin{equation}
\Delta t_{\rm obs}=\Delta t_{\rm int}+\Delta t_{\rm LIV}+\Delta t_{\rm spe}+\Delta t_{\rm DM}+\Delta t_{\rm gra}\ .
\end{equation}
Here $\Delta t_{\rm int}$ is the intrinsic (astrophysical) time delay between two test photons. It is hard to estimate the exact value of $\Delta t_{\rm int}$, since the inner workings of such events can be complicated and are model-dependent. Thus in our analysis we assume $\Delta t_{\rm int} = 0$ and an upper limit of time delay induced by EEP can be achieved.  $\Delta t_{\rm LIV}$ is the time delay from the Lorentz invariance violation. It is ignored in the following analysis, since current observations have already put very stringent limits on this term \citep[e.g., see][]{2013PhRvD..87l2001V}. $\Delta t_{\rm spe}$ is the potential time delay due to special-relativistic effects with non-zero photon rest mass. Modern experiments have showed that $\Delta t_{\rm spe}$ is negligible even when the energy of test photons are lower than radio band \citep[e.g., see][]{2007PPCF...49..429R}. $\Delta t_{\rm DM}$ represents the time delay contributed by the dispersion from the line-of-sight free electron content, which is non-negligible especially for low energy photons (e.g., radio signals). $\Delta t_{\rm gra}$ corresponds to the difference in arrival time of two photons of energy $E_1$ and $E_2$, caused by the gravitational potential $U(r)$ integrated from the emission source to Earth. From Equation (1) we can write
\begin{equation}
\Delta t_{\rm gra}=\frac{\gamma_{\rm 1}-\gamma_{\rm 2}}{c^3}\int_{r_o}^{r_e}~U(r)dr\;,
\end{equation}
where $U(r)$ can be decomposed into three components, that is, the gravitational fields of the host galaxy of the transient,  intergalactic background field, as well as Laniakea supercluster of galaxies in which the Milky Way galaxy as well as the Local Group reside. Since in the local universe the contribution from home supercluster can dominate the other two components,  in this analysis we only consider the contribution from the Laniakea $U_{L}(r)$. Thus we have
\begin{equation}
\Delta t_{\rm gra}=\frac{\gamma_{\rm 1}-\gamma_{\rm 2}}{c^3}\int_{r_o}^{r_e}~U_{L}(r)dr\;.
\end{equation}

Assuming $\Delta t_{\rm int}>0$ and casting off the negligible components, we have
\begin{equation}
\Delta t_{\rm obs}-\Delta t_{\rm DM}>\frac{\gamma_{\rm 1}-\gamma_{\rm 2}}{c^3}\int_{r_o}^{r_e}~U_{L}(r)dr\;.
\end{equation}
Although the gravitational potential of the Laniakea $U_{L}(r)$ at large distances is still not known, we still adopt the Keplerian potential $U_{L}(r)=-GM/r$ here. Thus we have \citep{1988PhRvL..60..173L}
\begin{equation}
%\begin{split}
\Delta t_{\rm obs}> \left(\gamma_{1}-\gamma_{2}\right ) \frac{GM_{\rm L}}{c^{3}} \times
\ln \left\{ \frac{ \left[d+\left(d^{2}-b^{2}\right)^{1/2}\right] \left[r_{L}+s_{\rm n}\left(r_{L}^{2}-b^{2}\right)^{1/2}\right] }{b^{2}} \right\}\;,
%\end{split}
\label{eq:gammadiff}
\end{equation}
where $G=6.68\times10^{-8}{\rm erg\cdot cm\cdot g}^{-2}$ is the gravitational constant, $M_{\rm L}\simeq1\times10^{17}M_{\odot}$ is the Laniakea mass \citep{2014Natur.513...71T}, $c=3\times10^{10}{\rm cm\;s^{-1}}$ is the energy-independent speed for massless particles, $d$ is the distance from the source to the Laniakea center (if the source is of cosmological origin, $d$ is approximated as the distance from the source to the Earth), $b$ is the impact parameter of the light rays relative to the Laniakea center, and $s_{\rm n}=\pm1$ represents the sign of the correction of the source direction.
If the source is located along the direction of Laniakea center, $s_{\rm n}=+1$. While, $s_{\rm n}=-1$ corresponds to the source located along the direction of anti-Laniakea center.
However, the center coordinates of Laniakea itself is not well estimated. Since the gravitational center of Laniakea is believed to be the so-called Great Attractor \citep{1988ApJ...326..19L}, a gravity anomaly in intergalactic space within the vicinity of the Hydra-Centaurus Supercluster that reveals the existence of a localised concentration of mass tens of thousands of times more massive than the Milky Way \citep{2014Natur.513...71T}, we consider the center of the Great Attractor instead (i.e., $\rm R.A.=10^{h}32^{m}$, $\rm decl.=-46^{\circ}00^{'}$).
For a cosmic source in the direction ($\rm R.A.=\beta_{s}$, $\rm decl.=\delta_{s}$), the impact parameter can be expressed as
 \begin{equation}
b=r_{L}\sqrt{1-(\sin \delta_{s} \sin \delta_{L}+\cos \delta_{s} \cos \delta_{L} \cos(\beta_{s}-\beta_{L}))^{2}}\;,
\end{equation}
where $r_{L}=79$ Mpc is the distance from the Earth to the Laniakea center, and ($\beta_{L}=10^{\rm h}32^{\rm m}$, $\delta_{L}=-46^{\circ}00^{'}$) are the coordinates of the Laniakea center in the equatorial coordinate system.

%It consists of four subparts which are Virgo Supercluster, Hydra-Centaurus Supercluster, Pavo-Indus Supercluster and Southern Supercluster. The Virgo  supercluster is only the subsidiary of the Laniakea, so the Milky Way located the peripheral zone. As a result of this, the Supercluster which is placed for the Milky Way expanded several times.
%
\section{Test the EEP with Cosmic Transient Events}
It can be seen from the previous section that the larger the distance of the transient, the shorter the time delay, the better the EEP constraint. Flares of TeV blazars, GRBs as well as FRBs are among common transients in the Universe. All of these events are thought to be extragalactic origins, and have short intrinsic variation time scales, thus providing ideal testbed for obtaining EEP constraints. In our analysis, we select some examples from each of these three groups.

As a subclass of active galactic nuclei, the blazars can be divided into flat spectrum radio quasars if they have strong emission lines and BL lacertae or not. The broadband non-thermal emission of blazars extend from radio up to high-energy and very-high-energy \citep{1997ARA&A..35..445U}. Because of their cosmological distances, fast variability as well as VHE photons emitted in the TeV band, such TeV blazars can be used to constrain EEP \citep{2016ApJ...818L...2W}. Since the TeV blazar PKS 2155-304 with a redshift of $z=0.117$ \citep{1974AuJPA..32....1S} lies beyond the realm of Laniakea, we can constrain EEP with time delay in Laniakea's gravitational field with this source. Its coordinates (J2000) are $\rm R.A.=21^{h}58^{m}52.6^{s}$, $\rm decl.=-30^{\circ}13^{'}18^{''}$. The time delay between 0.2-0.8 TeV and $>0.8$ TeV photons is 73 s \citep{2008PhRvL.101q0402A}. From Equation (6), we have
 \begin{equation}
\gamma_{(0.2TeV-0.8TeV)}-\gamma_{(>0.8TeV)}<5.1\times10^{-11}\;.
\end{equation}

GRBs were discovered by the Vela satellites in the late 1960¡¯s. As the most extreme catastrophic events, the energies of photons are mainly in the keV to MeV band. The duration of the prompt emission ranges from 0.1 s to 1000 s, which can be devided into short duration ($T < 2$ s) and long duration ($T > 2$ s) groups \citep[e.g., see][]{2015PhR...561....1K}. Using GRBs for EEP constraints has several major advantages, including unprecedented spectral coverage for seven orders of magnitude in energy, the easy identification of arrival time lag between different energy photons, as well as the potential to determine the arrival time lag between $\gamma$-ray photons to lower energy photons. In this analysis we take two GRBs, GRB 090510 and GRB 080319B, as our samples, the same as \cite{2015ApJ...810..121G}.

The short burst GRB 090510 was detected by the Fermi Gamma-ray Space Telescope, with coordinates $\rm R.A.=22^{h}14^{m}12^{s}.47$, $\rm decl.=-26^{\circ}35^{'}00^{''}.4$ \citep{2009GCNR..218....1H}. Its redshift is $z=0.903\pm0.003$ \citep{2009GCN..9353....1R,2010A&A...516A..71M}. The nominal time delay between the GeV photons and MeV photons is $\bigtriangleup t\simeq0.83s$ (Abdo et al. 2009). While GRB 080319B was detected by Swift satellite with coordinates (J2000) $\rm R.A.=14^{h}31^{m}40^{s}.7$, $\rm decl.=+36^{\circ}18^{'}14^{''}.7$ \citep{2008Natur.455..183R}. The redshit is $z=0.937$ \citep{2008GCN..7451....1V}. The longest time delay between the MeV photons and the photons of the optical flash has a value of 5 s.  Thus from Equation (6) we have

(1) $\gamma_{GeV}-\gamma_{MeV}<3.1\times10^{-13}$ for GRB 090510,

(2) $\gamma_{eV}-\gamma_{MeV}<1.9\times10^{-12}$ for GRB 080319B.

FRB is a new type of millisecond radio burst transients which has caused widespread concern. The first detected FRB named FRB 010724 was found in a search for pulsars using a technique to detect bright single pulses with the Parkes radio telescope \citep{2007Sci...318..777L}. Following this, a series of FRBs have been reported, with a present total of over fifteen bursts. Most of these FRBs have high dispersion measures (DM), especially for FRB 121002 which is up to 1629.18 $\rm pc\cdot cm^{-3}$ \citep{2015arXiv151107746C}. Because of the high DM, many researchers believe that FRBs are of cosmological origin \citep[e.g., see][]{2013Sci...341...53T}. If FRBs are indeed cosmological events, they can provide more stringent constraints on the EEP, with shorter time delay between different frequencies. Moreover, it seems that several FRBs may be associated with GRBs \citep{2014ApJ...783..L35G}. If proved to be true, such associations can also provide an interesting tool for EEP constraints. We take three FRB examples here, FRB 110220, FRB/GRB 101011A and FRB/GRB 100704A, the same as \citep{2015PhRvL.115z1101W}.

FRB 110220 was discovered by the 64-m Parkes multibeam radio telescope with coordinates (J2000) $\rm R.A.=22^{h}34^{m}$, $\rm Dec.=-12^{o}24^{'}$ \citep{2013Sci...341...53T}. From the DM value, Its redshift is inferred to be $z\approx 0.81$. The arrival time delay is identified to be $\Delta t\simeq1s$ for photons ranging in frequency from abut 1.5 GHz to 1.2 GHz. While FRB/GRBs 101011A and 100704A are two possible FRB/GRB associations \citep{2014ApJ...783..L35G}. FRB/GRB 101011A was detected and located by Swift/BAT with coordinates (J2000) $\rm R.A.=03^{h}13^{m}12^{s}$, $\rm Dec.=-65^{o}59^{'}08^{''}$ \citep{2010GCN..11331...1C}. GRB/FRB 100704A was also detected by Swift/BAT with coordinates (J2000) $\rm R.A.=08^{h}54^{m}33^{s}$, $\rm Dec.=-24^{o}12^{'}55^{''}$ \citep{2010GCN..10929...1G}. The estimated values of the redshifts of these associations are $z\geq0.246$ for GRB/FRB 101011A and $z\geq0.166$ for GRB/FRB 100704A from the Amati relation \citep{2014ApJ...783..L35G}. The time delays between different radio frequencies are $\Delta t\simeq0.438$s for GRB/FRB 101011A and $\Delta t\simeq0.149$s for GRB/FRB 100704A \citep{2012ApJ...757...38B}. From Equation (6), we have

(1) $\gamma_{1.2GHz}-\gamma_{1.5GHz}<3.6\times10^{-13}$ for FRB 110220,

(2) $\gamma_{1.23GHz}-\gamma_{1.45GHz}<2.1\times10^{-13}$ for FRB/GRB 101011A,

(3) $\gamma_{1.23GHz}-\gamma_{1.45GHz}<5.9\times10^{-14}$ for FRB/GRB 100704A.

\section{Summary and Discussion}
\label{sect:Conclusion}
It can be seen in the previous section that, by applying a supercluster (instead of Milky Way) as the source of gravitational field, we obtain a constraint on the accuracy of the EEP 4--5 orders of magnitudes more stringent than any previously analysis. Furthermore, we obtain the EEP constraints by using various types of cosmic transients with different spectral energy distributions at different redshift ranges.
Although with the Laniakea supercluster of galaxies a more stringent EEP constraint can be obtained, this is still a conservative upper limit. We neglected the other potential contributions to the observed time delay $\Delta t_{\rm obs}$ (see Equation~2) in these calculations. In reality, $\Delta t_{\rm grav}$ could be much shorter than $\Delta t_{\rm obs}$. The inclusion of contributions from these neglected components in $\Delta t_{\rm obs}$ could further improve these limits on EEP.

In this work, the coordinates and the gravitational field of the Laniakea supercluster bring most uncertainties. The coordinates we adopted ($\rm R.A.=\beta_{s}=10^{h}32^{m}$, $\rm decl.=\delta_{s}=-46^{\circ}00^{'}$) are the center coordinates of the Great Attractor, which is the central region of the Laniakea. However, it is quite possible that the center of the Laniakea and the Great Attractor are not aligned with each other. Fortunately, this uncertainty can only cause a very small effect on our results. For one thing it can be seen from Equation (6) that the effect of inaccurate coordinates is small to the final results. Besides we can try to constrain EEP with the gravity of Great Attractor only. The mass of the Great Attractor is $1\times10^{16}$ solar masses. If we adopt the gravitational potential of the Great Attractor for EEP constraints, we have $\gamma_{(0.2TeV-0.8TeV)}-\gamma_{(>0.8TeV)}<5.1\times10^{-10}$ for PKS 2155-304, $\gamma_{GeV}-\gamma_{MeV}<3.1\times10^{-12}$ for GRB 090510, $\gamma_{eV}-\gamma_{MeV}<1.9\times10^{-11}$ for GRB 080319B, $\gamma_{1.2GHz}-\gamma_{1.5GHz}<3.6\times10^{-12}$ for FRB 110220, $\gamma_{1.23GHz}-\gamma_{1.45GHz}<2.1\times10^{-12}$ for FRB/GRB 101011A, and $\gamma_{1.23GHz}-\gamma_{1.45GHz}<5.9\times10^{-13}$ for FRB/GRB 100704A. It leads to EEP constraints one order of magnitude worse than those constraints presented in Section 3, and still 3-4 orders of magnitude better than previous works. However, since the Milky Way does not lie within the Great Attractor, such EEP constraints with gravitational potential of the Great Attractor cannot be applied to every transient in the sky.

The gravitational field of Laniakea supercluster is a complicated issue. As mentioned above, such structures are not bound by self-gravity. That is, a simple Keplerian or NFW potential \citep{1996ApJ...462..563N} may not well describe the whole system. The exact mass distribution within Laniakea is still uncertain till today, and we also calculate the constraints with a single NFW as well as isothermal potentials.
Generally speaking, the results from Keplerian and isothermal potentials are nearly the same
\citep[see also][]{1988PhRvL..60..176K}. While the NFW potential can bring extra uncertainties,
as the scale radius $r_s$ of the gravitational source is unknown. However even in the worst
situation the constraints from the NFW potential are compatible with those from the Keplerian
potential within 1 order of magnitude \citep[see also][]{2015arXiv160104558Z}. Thus,
it seems reasonable to assume a simple form of the gravitational potential (i.e. the Keplerian
potential) for the whole supercluser.

Also the intrinsic time delay $\Delta t_{\rm int}$ may bring extra uncertainties to our analysis.
With the assumption that $\Delta t_{\rm int}\geq0$, the conservative upper limits on
the EEP can be obtained by adopting $\Delta t_{\rm int}=0$ for all of our samples.
However, in the chance of $\Delta t_{\rm int}$ being negative, the real upper limit
would be larger than the ``upper limit" derived from the assumption of $\Delta t_{\rm int}=0$.
Fortunately, most of the intrinsic time delays of our samples are positive.
GRB 090510 shows an observed delay $\simeq 0.83$ s of GeV photon \citep{2009Natur.462..331A}.
Since GeV emissions from GRBs are considered as inverse Compton scattering of MeV
or lower energy photons, such components should arise later intrinsically, i.e.,
$\Delta t_{\rm int}>0$. Similarly, the $>0.8$ TeV signals from PKS 2155-304 arrive
later than lower energy ones, which is compatible with the prediction that
high energy photons should be emitted later in the framework of inverse Compton scattering.
Since the radio emissions of FRBs are non-thermal, cooling of electrons should
give rise to high frequency emissions earlier than lower frequency components.
However, the intrinsic time delay for GRB 080319B may be negative. The MeV component
of GRB 080319B can be interpreted as inverse Compton scattering of optical emissions,
thus maybe arise later intrinsically, while the MeV photons observationally arrive
earlier than the optical photons.

Considering that $\Delta t_{\rm LIV}$, $\Delta t_{\rm spe}$, and $\Delta t_{\rm DM}$
in Equation~(2) are negligible for the analysis of this work, and assuming $\Delta t_{\rm int}<0$
for GRB 080319B, one can derive $|\Delta t_{\rm obs}|=|\Delta t_{\rm gra}|-|\Delta t_{\rm int}|$,
where $|\Delta t_{\rm obs}|=5$ s. To account for the uncertainty of the intrinsic time delay,
we also test two more cases by assuming $|\Delta t_{\rm int}|\leq|\Delta t_{\rm obs}|$
and $|\Delta t_{\rm int}|\gg|\Delta t_{\rm obs}|$. (i) For the case of $|\Delta t_{\rm int}|\leq|\Delta t_{\rm obs}|$,
we find that the implications of GRB 080319B tests of the EEP are not greatly affected, i.e.,
the constraint results vary within a factor of two.
(ii) In principle, the value of the travel time delay ($|\Delta t_{\rm gra}|$) caused by the gravitational
potential is small. However, for the case of $|\Delta t_{\rm int}|\gg|\Delta t_{\rm obs}|$,
$|\Delta t_{\rm gra}|$ has always to be on the same order of $|\Delta t_{\rm int}|$
(i.e., $|\Delta t_{\rm gra}|=|\Delta t_{\rm obs}|+|\Delta t_{\rm int}|\approx|\Delta t_{\rm int}|$)
to make $|\Delta t_{\rm obs}|$ be 5 s for taking different $|\Delta t_{\rm int}|$. Note that $\Delta t_{\rm int}$ is the intrinsic time delay by the emission source, while $\Delta t_{\rm gra}$ is the external time delay caused by the traveling in an external gravitational field. Hence,
it should be a coincidence at a high confidence level for this case of $|\Delta t_{\rm int}|\gg|\Delta t_{\rm obs}|$.

Above content provides some new ideas. From Equation (6), it shows that there are much better results with larger mass of the galaxy supercluster bringing the gravitational potential and greater distances of the cosmic transients to the Milky Way. If the galaxy supercluster is fixed, one can look for acceptable cosmic transients according to the coordinates of the Milky Way and the galaxy supercluster.

\section*{Acknowledgments}
We are grateful to the anonymous referee for useful suggestions and comments.
This work is partially supported by the National Basic Research Program (``973'' Program)
of China (Grants 2014CB845800 and 2013CB834900), the National Natural Science Foundation
of China (grants Nos. 11322328 and 11433009),
the Youth Innovation Promotion Association (2011231), and the Strategic Priority Research Program
``The Emergence of Cosmological Structures'' (Grant No. XDB09000000) of
the Chinese Academy of Sciences.

%
%\bibliographystyle{model1a-num-names}

\begin{thebibliography}{}


\bibitem[Abdo et al.(2009)]{2009Natur.462..331A} Abdo, A. A.,
Ackerman, M, Ajello, M., et al.\ 2009, Nature, 462, 331

\bibitem[Aharonian et al.(2008)]{2008PhRvL.101q0402A} Aharonian, F.,
Akhperjanian, A.~G., Barres de Almeida, U., et al.\ 2008, Physical Review
Letters, 101, 170402

\bibitem[Bannister et al.(2012)]{2012ApJ...757...38B} Bannister, K.~W.,
Murphy, T., Gaensler, B.~M., \& Reynolds, J.~E.\ 2012, ApJ, 757, 38


\bibitem[Bertotti et al.(2003)]{2003Natur.425..374B} Bertotti, B., Iess,
L., \& Tortora, P.\ 2003, Nature, 425, 374

\bibitem[Cannizzo et al.(2010)]{2010GCN..11331...1C} Cannizzo, J.~K.,
Barthelmy, S.~D., Baumgartner, W.~H., et al.\ 2010, GRB Coordinates
Network, 11331, 1

\bibitem[Champion et al.(2015)]{2015arXiv151107746C} Champion, D.~J.,
Petroff, E., Kramer, M., et al.\ 2015, arXiv:1511.07746

\bibitem[Deng \& Zhang (2014)]{2014ApJ...783..L35G} Deng, W., Zhang, B.\ 2014, ApJ, 783, L35


\bibitem[Gao et al.(2015)]{2015ApJ...810..121G} Gao, H., Wu, X.-F.,
\& M{\'e}sz{\'a}ros, P.\ 2015, ApJ, 810, 121


\bibitem[Grupe et al.(2010)]{2010GCN..10929...1G} Grupe, D., Barthelmy,
S.~D., Cummings, J.~R., et al.\ 2010, GRB Coordinates Network, 10929, 1


\bibitem[Hoversten et al.(2009)]{2009GCNR..218....1H} Hoversten, E.~A.,
Krimm, H.~A., Grupe, D., et al.\ 2009, GCN Report, 218, 1


\bibitem[Kahya
\& Desai(2016)]{2016arXiv160204779K} Kahya, E.~O., \& Desai, S.\ 2016, arXiv:1602.04779


\bibitem[Krauss
\& Tremaine(1988)]{1988PhRvL..60..176K} Krauss, L.~M., \& Tremaine, S.\ 1988, Physical Review Letters, 60, 176


\bibitem[Kumar
\& Zhang(2015)]{2015PhR...561....1K} Kumar, P., \& Zhang, B.\ 2015, Phys. Rep., 561, 1


\bibitem[Lambert
\& Le Poncin-Lafitte(2011)]{2011A&A...529A..70L} Lambert, S.~B., \& Le Poncin-Lafitte, C.\ 2011, A\&A, 529, A70


\bibitem[Lambert
\& Le Poncin-Lafitte(2009)]{2009A&A...499..331L} Lambert, S.~B., \& Le Poncin-Lafitte, C.\ 2009, A\&A, 499, 331


\bibitem[Longo(1988)]{1988PhRvL..60..173L} Longo, M.~J.\ 1988, Physical
Review Letters, 60, 173


\bibitem[Lorimer et al.(2007)]{2007Sci...318..777L} Lorimer, D.~R., Bailes,
M., McLaughlin, M.~A., Narkevic, D.~J.,
\& Crawford, F.\ 2007, Science, 318, 777


\bibitem[Lynden-Bell et al.(1988)]{1988ApJ...326..19L} Lynden-Bell, D., Faber,
S. M., Burstein, D., et al.\ 1988, ApJ, 326, 19


\bibitem[McBreen et
al.(2010)]{2010A&A...516A..71M} McBreen, S., Kr{\"u}hler, T., Rau, A., et al.\ 2010, A\&A, 516, A71


\bibitem[Navarro et al.(1996)]{1996ApJ...462..563N} Navarro, J. F, Frenk, C. S.,
\& White, S. D. M.\ 1996, ApJ, 462, 563


\bibitem[Nusser(2016)]{2015arXiv160103636N} Nusser, A. \ 2016, arXiv:1601.03636


\bibitem[Racusin et al.(2008)]{2008Natur.455..183R} Racusin, J.~L., Karpov,
S.~V., Sokolowski, M., et al.\ 2008, Nature, 455, 183


\bibitem[Rau et al.(2009)]{2009GCN..9353....1R} Rau, A., McBreen, S.,
\& Kruehler, T.\ 2009, GRB Coordinates Network, 9353, 1


\bibitem[Ryutov(2007)]{2007PPCF...49..429R} Ryutov, D.~D.\ 2007, Plasma
Physics and Controlled Fusion, 49, B429


\bibitem[Shapiro(1964)]{1964PhRvL..13..789S} Shapiro, I.~I.\ 1964, Physical
Review Letters, 13, 789


\bibitem[Shimmins
\& Bolton(1974)]{1974AuJPA..32....1S} Shimmins, A.~J., \& Bolton, J.~G.\ 1974, Australian Journal of Physics Astrophysical Supplement, 32, 1


\bibitem[Thornton et al.(2013)]{2013Sci...341...53T} Thornton, D.,
Stappers, B., Bailes, M., et al.\ 2013, Science, 341, 53


\bibitem[Tingay
\& Kaplan(2016)]{2016arXiv160207643} Tingay, S.~J., \& Kaplan, D.~L.\ 2016, arXiv:1602.07643


\bibitem[Tully et al.(2014)]{2014Natur.513...71T} Tully, R.~B., Courtois,
H., Hoffman, Y., \& Pomar{\`e}de, D.\ 2014, Nature, 513, 71


\bibitem[Ulrich et
al.(1997)]{1997ARA&A..35..445U} Ulrich, M.-H., Maraschi, L., \& Urry, C.~M.\ 1997, ARA\&A, 35, 445


\bibitem[Vasileiou et al.(2013)]{2013PhRvD..87l2001V} Vasileiou, V.,
Jacholkowska, A., Piron, F., et al.\ 2013, Physical Review D, 87, 122001


\bibitem[Vreeswijk et al.(2008)]{2008GCN..7451....1V} Vreeswijk, P.~M.,
Milvang-Jensen, B., Smette, A., et al.\ 2008, GRB Coordinates Network,
7451, 1


\bibitem[Wang et al.(2016)]{2016arXiv160206805W} Wang, Z.-Y., Liu, R.-Y.,
\& Wang, X.-Y.\ 2016, arXiv:1602.06805


\bibitem[Wei et al.(2015)]{2015PhRvL.115z1101W} Wei, J.-J., Gao, H., Wu,
X.-F., \& M{\'e}sz{\'a}ros, P.\ 2015, Physical Review Letters, 115, 261101


\bibitem[Wei et al.(2016)]{2016ApJ...818L...2W} Wei, J.-J., Wang, J.-S.,
Gao, H., \& Wu, X.-F.\ 2016, ApJ, 818, L2


\bibitem[Will(2014)]{2014LRR....17....4W} Will, C.~M.\ 2014, Living Reviews
in Relativity, 17,


\bibitem[Will(2006)]{2006LRR.....9....3W} Will, C.~M.\ 2006, Living Reviews
in Relativity, 9,


\bibitem[Will(1993)]{1993tegp.book.....W} Will, C.~M.\ 1993, Theory and
Experiment in Gravitational Physics, by Clifford M.~Will, pp.~396.~ISBN
0521439736.~Cambridge, UK: Cambridge University Press, March 1993., 396


\bibitem[Wu et al.(2016)]{2016arXiv160201566W} Wu, X.-F., Gao, H., Wei,
J.-J., et al.\ 2016, arXiv:1602.01566


\bibitem[Zhang(2016)]{2015arXiv160104558Z} Zhang, S.~N.\ 2016, arXiv:1601:04558




\end{thebibliography}

%
\end{document}